\documentclass[11pt,DIV=12,a4paper,BCOR=0mm,numbers=noenddot]{scrartcl} 
\usepackage{amsmath,amssymb}
\usepackage{graphicx}
\usepackage{color}
\usepackage[utf8]{inputenc}

\definecolor{darkblue}{rgb}{0.0,0.0,0.4}
\definecolor{darkgreen}{rgb}{0.0,0.4,0.0}
\usepackage[colorlinks=true]{hyperref}
\hypersetup{linkcolor=darkblue, citecolor=darkgreen, urlcolor=darkblue}
\usepackage{doi}
\usepackage[sort&compress,numbers]{natbib}

\usepackage[T1]{fontenc}
\usepackage{textcomp}
\usepackage{upgreek}
\usepackage{fourier}
\usepackage[sfdefault, scaled=0.96]{overlock}

\usepackage{microtype}
\usepackage{mathtools}
\usepackage{siunitx}

\usepackage{caption} 
\usepackage{booktabs}

\renewcommand{\uppi}{\otherpi}
\newcommand{\dx}{\mathrm{d}}
\newcommand{\HI}{H_\mathrm{I}}
\newcommand{\HII}{H_\mathrm{II}}
\newcommand{\AI}{a_\mathrm{I}}
\newcommand{\AII}{a_\mathrm{II}}
\newcommand{\NI}{N_\mathrm{I}}
\newcommand{\NII}{N_\mathrm{II}}
\newcommand{\tII}{t_\mathrm{II}}
\newcommand{\mrm}[2]{#1_\mathrm{#2}}
\newcommand{\mpl}{\mrm{m}{Pl}}
\def\lsim{\mathrel{\rlap{\lower4pt\hbox{$\sim$}}
		\raise1pt\hbox{$<$}}}                
\def\gsim{\mathrel{\rlap{\lower4pt\hbox{$\sim$}}
		\raise1pt\hbox{$>$}}}                

\renewcommand\({\left(}
\renewcommand\){\right)}
\renewcommand\[{\left[}
\renewcommand\]{\right]}

\begin{document}
\numberwithin{equation}{section}
\title{
\vspace{2.0cm} 
\LARGE{\textbf{QCD axions and axionlike particles in a two-inflation scenario
\vspace{0.3cm}}}}

\linespread{1.0}\selectfont

\author{{\large Sebastian Hoof$^{1}$ and Joerg Jaeckel$^{2}$}\\[2ex]
	\small \em $^1$Department of Physics, Imperial College London,\\
	\small \em Blackett Laboratory, Prince Consort Road, London SW7 2AZ, United Kingdom\\
	\small \href{mailto:s.hoof15@imperial.ac.uk}{s.hoof15@imperial.ac.uk}\\
	\small \em $^2$Institut f\"ur Theoretische Physik, Ruprecht-Karls-Universit\"at Heidelberg,\\
	\small \em Philosophenweg 16, 69120 Heidelberg, Germany\\
	\small \href{mailto:jjaeckel@thphys.uni-heidelberg.de}{jjaeckel@thphys.uni-heidelberg.de}}

\date{}
\clearpage\maketitle
\thispagestyle{empty}

\linespread{1.3}\selectfont

\begin{abstract}
\noindent \textbf{\textsf{Abstract:}} We investigate the phenomenology of QCD axions and axionlike particles in a scenario with two eras of inflation. In particular, we describe the possible solutions for the QCD axion field equation after the second inflation and reheating. We calculate the dilution numerically for QCD axions and give an analytic approximation for axionlike particles. While it has been realised before that such a scenario can dilute the axion energy density and open up the parameter space for the axion decay constant~$f_A$, we find that even a small increase in the relative QCD axion energy density is possible. 
\end{abstract}

\newpage
\section{Introduction}\label{sec:intro}
Axions first appeared as a solution for the strong CP problem via the Peccei-Quinn~(PQ) mechanism~\cite{Peccei:1977hh, Peccei:1977ur}. Weinberg~\cite{Weinberg:1977ma} and Wilczek~\cite{Wilczek:1977pj} then realised that this mechanism gives rise to a new pseudoscalar particle, the QCD axion. Experimental evidence against the initial axion models linked to the weak scale led to the introduction of ``invisible axions'', which turned out to be excellent dark matter candidates~\cite{Preskill:1982cy, Abbott:1982af, Dine:1982ah, Turner:1985si}. The initial concept of a QCD axion was also generalised to axionlike particles~(ALPs), which can be theoretically motivated by beyond-the-Standard-Model physics such as String theory~\cite{Witten:1984dg, hep-th/0605206,Cicoli:2012sz} (see e.g.~\cite{Kim:1986ax,Jaeckel:2010ni} for more general reviews). For our purposes, an ALP is a pseudoscalar particle with a fundamental shift symmetry and a not explicitly temperature-dependent mass.

Axion~\cite{Preskill:1982cy, Abbott:1982af, Dine:1982ah, Turner:1985si} and ALP~\cite{1201.5902} cold dark matter can be produced non-thermally by the so-called misalignment mechanism, i.e. from oscillations of an axion field that initially is not in the minimum of its potential. The axion energy density from this mechanism can be calculated numerically and various approximations of the result are given in the literature~\cite{Turner:1985si, Lyth:1991ub, astro-ph/9405028, Bae:2008ue, 0903.4377,0910.1066,1412.0789}. We will work with numerical solutions and semi-analytic approximations, but it is useful to have an expression for back-of-the-envelope calculations. Here, we quote a simplified version of~\cite{0910.1066},
\begin{equation}
   \Omega_{A,o}^\mathrm{std}h^2 \sim \num{0.09} \; \mrm{\theta}{i}^2 \left(\frac{76}{\mrm{g}{osc}}\right)^{0.41} \left(\frac{f_A}{\SI{e12}{\giga\electronvolt}}\right)^{1.19} \, , \label{eq:relicdensityapprox}
\end{equation}
where $f_A$ is the QCD axion decay constant and $\mrm{g}{osc}$ are the effective relativistic degrees of freedom when the axion starts to become dynamical. If the PQ symmetry breaking occurs before inflation, the initial angle $\mrm{\theta}{i}$ is essentially a free parameter. On the other hand, if the PQ symmetry breaks after inflation, the QCD axion energy density today is entirely determined by the PQ scale~$f_A$ as long as we assume standard cosmological evolution. This is because the axion field value at the Peccei-Quinn transition is chosen independently in causally disconnected regions. Today's universe contains a very large number of these regions and the axion energy density is fixed by an average. This sets a limit on the axion mass~\cite{Turner:1985si}.
In this scenario topological defects, in particular strings, may also contribute significantly to today's axion density (although the precise amount is somewhat uncertain)~\cite{Sikivie:2006ni,Hiramatsu:2012gg,Kawasaki:2014sqa,Fleury:2015aca}.

Using relation~(\ref{eq:relicdensityapprox}), the critical energy density of QCD axions today would exceed the observed amount of dark matter, $\Omega_{c,o}h^2 \approx 0.12$~\cite{1502.01589}, for $f_A\gsim\SI{E12}{\giga\electronvolt}$ and $\mrm{\theta}{i}\sim 1$. It has been noted that inflationary physics can avoid this bound on $f_A$ by further diluting the axion energy density. This can happen during primordial inflation~\cite{Nanopoulos:1983cf}, late inflation~\cite{Dimopoulos:1988pw} or, as we do, by introducing a second inflationary era~\cite{1507.08660}. Other mechanisms to dilute the axion energy density are entropy dilution~\cite{Dine:1982ah, Steinhardt:1983ia, Yamamoto:1985mb, Lazarides:1987zf} or hidden magnetic monopoles~\cite{1511.05030}. Axion have furthermore been studied in non-standard cosmologies such as low temperature reheating or kination cosmology~\cite{0912.0015}.

In this work, we perform a detailed investigation of the effects of a second stage of inflation. We mainly focus on the case of a homogeneous axion or ALP field as one would expect if the axion or ALP is already present during inflation. We only briefly comment on the case with strong inhomogeneities and topological defects that one would expect, e.g., in a scenario where the PQ phase transition only occurs after the (first stage) of inflation. This interesting case is left for future work.

For axionlike particles and for QCD axions in certain parameter ranges, we find the {na\"ively} expected dilution. However, reheating the Universe after the second period of inflation can cause deviations from this effect for QCD axions in a significant parameter range. Particularly interesting, non-trivial behaviour occurs if the axion field oscillations start before the end of the second inflation and if the reheating temperature is higher than the QCD scale. Beyond axions, our findings may generally apply to models with an explicitly temperature-dependent mass. 

This paper is organised as follows: Next, in Sec.~\ref{sec:2flation}, we introduce a simplistic description of two-inflation, a scenario with two episodes of accelerated expansion of the early Universe. In the following section, we describe how the axion energy density is affected by this scenario and we derive an analytic expression for ALPs as well as numerical solutions for QCD axions. The findings and interesting phenomenological aspects are discussed in Sec.~\ref{sec:findings1} for ALPs and Sec.~\ref{sec:findings2} for QCD axions. We conclude with a discussion and outlook in Sec.~\ref{sec:discussion}.

\section{A simplistic two-inflation scenario}\label{sec:2flation}
While there exists a vast landscape of viable inflationary models (cf. e.g.~\cite{1303.3787,1502.02114}), we want to focus on generic features and consider a simplistic description of the Hubble parameter in the early Universe. 

\subsection{Evolution of the Hubble parameter}
The inflationary eras are realised by two additional cosmological constants, $H_\mathrm{I}^2 \gg H_\mathrm{II}^2$, which are ``turned off'' at some point to instantly turn into relativistic degrees of freedom and reheat the Universe.\footnote{Under certain conditions and for several models of inflation, an instantaneous transition may be problematic~\cite{1504.03768}.} This happens at scale factors $\AI$ and $\AII$, where fractions of the inflation energy densities are converted into relativistic degrees of freedom:
\begin{equation}
   H^2(a) = \begin{dcases}
   \HI^2 + \HII^2 \hfil & \text{for } a_\mathrm{ini} \leq  a <  \AI\\
   \HI^2 \left( \frac{\AI}{a}\right )^4 + \HII^2 & \text{for } \AI \leq  a < \AII\\
   \HI^2 \left( \frac{\AI}{a}\right )^4 + \HII^2 \left( \frac{\AII}{a}\right )^4 & \text{for } \AII \leq a \ll \mrm{a}{MR}
   \end{dcases} \, , \label{eq:hubbleexpr}
\end{equation}
where $\mrm{a}{MR}$ is the scale factor at matter-radiation equality and $\mrm{a}{ini}$ is the scale factor at which the first inflation starts.\footnote{Later in Sec.~\ref{realistic} we turn to a slightly more realistic model where the conversion of the energy in the inflaton into radiation requires a finite time. Then Eq.~\eqref{eq:hubbleexpr} is modified.} 
Note that the first term in the second line neglects the fact that the effective relativistic degrees of freedom~$g$ change as a function of temperature or scale factor.\footnote{Taking the change in degrees of freedom into account, the result is corrected by a factor $\sim g\times g_{S}^{-4/3}$. This is usually a slowly changing function in time with values of order 1.} The number of e-folds of the first inflationary period can be defined as $\exp(\NI)\equiv \AI/\mrm{a}{ini}$ since $H_\mathrm{I} \gg H_\mathrm{II}$. For the second episode of inflation, however, there is no unique definition of the number of e-folds~$\NII$. Nonetheless, it seems sensible to define the start of the second period of inflation at the point~$a_q$ when the radiation and inflation energy densities are equal. Assuming that the only other relevant energy density is radiation, one can also show that the effective equation of state at this equilibrium point is $w_\mathrm{eff} = -1/3$. This corresponds to the boundary to an era of accelerated expansion (second Friedmann equation). Let us therefore define $\NII$ with respect to that starting point such that~$\exp(\NII)\equiv \AII/a_q$. The scale factor~$a_q$ can be calculated by equating radiation and inflation energy densities between $\AI$ and $\AII$:
\begin{equation}
   \HI^2 \left( \frac{\AI}{a_q} \right)^4 = \HII^2 \, . \label{eq:aqdef}
\end{equation}
\begin{figure}
	\centering
	\includegraphics[width=0.618\linewidth]{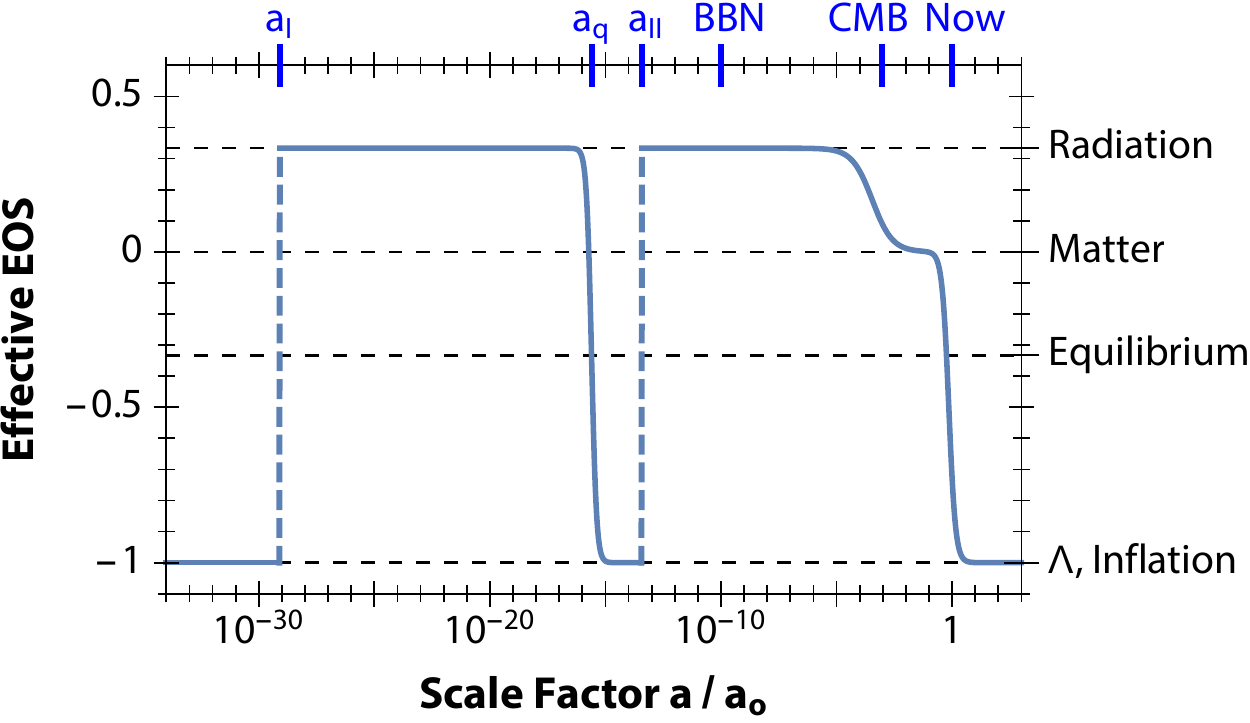}
	\caption[Cosmic history.]{Schematic overview of the cosmic history in the two-inflation scenario via the effective equation of state~(EOS). The scale factor is given as a fraction of the scale factor today and the labels on the right indicate the effectively dominating component of the energy density of the Universe. Equilibrium refers to the equality of radiation and inflation energy densities. Labels on the top indicate important events discussed in the text in addition to CMB formation~(CMB) and Big Bang nucleosynthesis~(BBN). \label{fig:inflationoverview}}
\end{figure}
We can eliminate the dependence on~$\AI$ in~(\ref{eq:hubbleexpr}) by using~(\ref{eq:aqdef}) such that
\begin{equation}
   H^2(a) =
   \begin{dcases}
   \HII^2 \left[ 1 + \mathrm{e}^{-4\NII} \left( \frac{\AII}{a}\right )^4 \right] & \text{for } \AI \leq  a < \AII\\
   \HII^2 \left[ 1 + \mathrm{e}^{-4\NII}\right] \left( \frac{\AII}{a} \right)^4 & \text{for } \AII \leq a \ll \mrm{a}{MR}
   \end{dcases} \, . \label{eq:Hcases}
\end{equation}
An overview of the cosmic history of such a scenario can be seen in Fig.~\ref{fig:inflationoverview}.

\subsection{The two-inflation parameter space}
Regarding the scales of $\HI$ and $\HII$, one can use several arguments to constrain them. First, we demand that the second reheating does not interfere with nucleosynthesis. This leads to $\mrm{T}{II,reh} \gsim \SI{10}{\mega \electronvolt}$ or, equivalently, $\HII \gsim \SI{4.5E-23}{\giga \electronvolt}$ from~(\ref{eq:TrehII}), neglecting the $\NII$-term. This is in line with studies that found a lower limit of a few~MeV for the reheating temperature~\cite{astro-ph/9811437, astro-ph/0002127, astro-ph/0403291}.

Placing an upper bound on the inflationary scales is, in principle, also possible via limits on the tensor-to-scalar ratio~$r_{0.05}$. A joint analysis of data from the BICEP2 and Keck Array Collaborations reports a 95\%-confidence level limit of~$r_{0.05}<0.09$~\cite{1510.09217} or $r_{0.05}<0.07$ including Planck results~\cite{1502.01589,1510.09217}. Using the definition of $r_{0.05}$ for slow-roll inflation, this can be turned into a limit of $\HII \lsim \SI{4E14}{\giga \electronvolt}$. 

For the present study the phenomenologically interesting, i.e. non-trivial, cases are limited to a regime where $\HII \lsim m_A(\mrm{T}{II})/3$, which is a much stronger restriction. The reason for this condition is that the axion field starts to oscillate around the time when $3H = m_A$, i.e. Hubble damping starts to become small.\footnote{This is the most commonly chosen point chosen for the start of the field oscillations, but others exist. The proportionality factor is not important for our case because we solve the relevant equations numerically around this point. See the discussion by Marsh for more details on this issue~\cite[sec.~4.3]{1510.07633}.} If we do not require this condition, the axion field does not start to oscillate before the end of the second inflation and the result would be no different to standard cosmology.\footnote{This statement is true for the axion energy density from the misalignment mechanism in a scenario where the axion is present during inflation, but is not necessarily the case for post-inflation PQ symmetry breaking and in particular for topological defects.}

\begin{figure}
	\centering
	\includegraphics[width=0.618\linewidth]{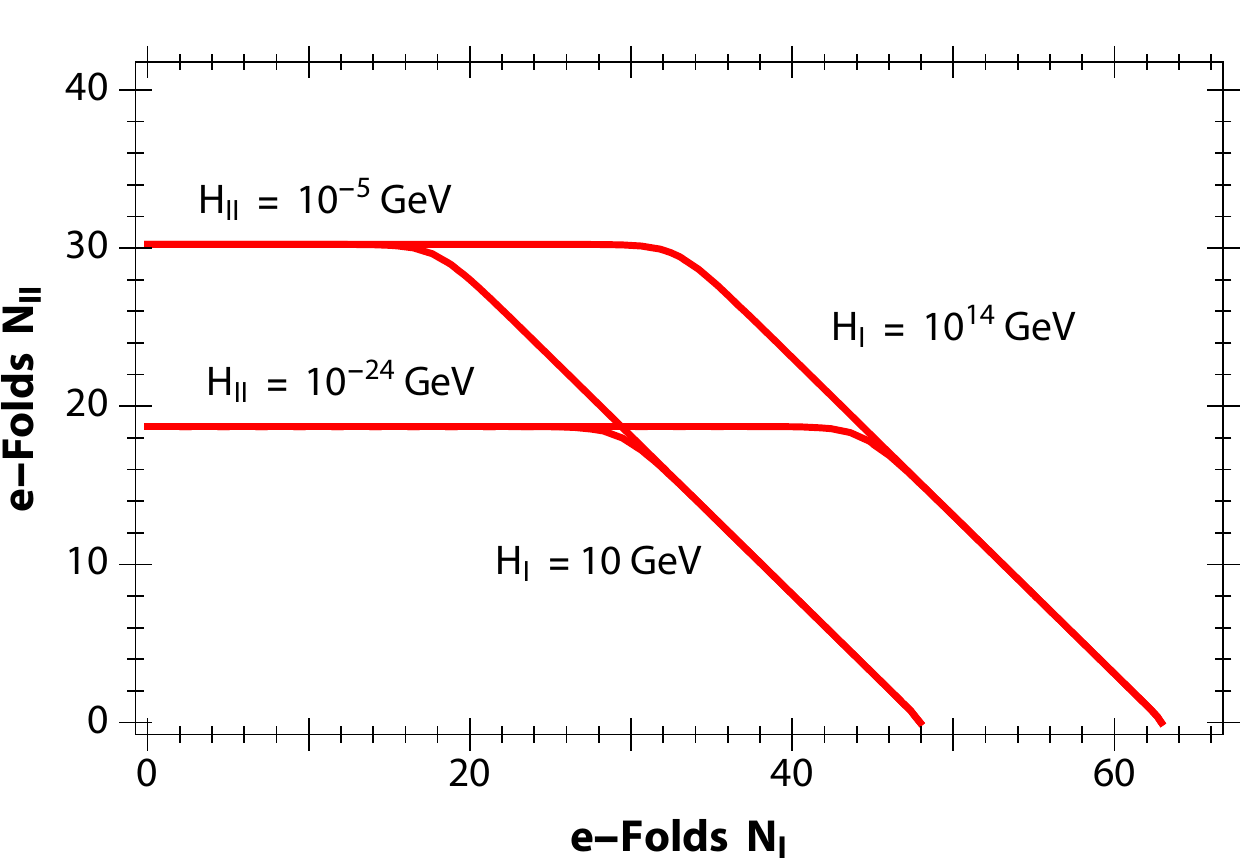}
	\caption[Horizon constraints on~$\NI$ and~$\NII$.]{Constraints on~$\NI$ and~$\NII$ from the horizon problem. The red lines indicate the minimal amount of inflation necessary to solve the horizon problem. Any point below those lines does not provide enough inflation to bring the Universe in causal contact. Having more e-folds in either direction is possible, but not necessary to solve the horizon problem.\label{fig:HorizonConstraints}}
\end{figure}

Note that the parameters of the inflationary periods are subject to constraints if we want inflation to solve the horizon and flatness problems~\cite{Guth:1980zm, Linde:1981mu}. The calculations for one era of inflation can be readily generalised to multiple inflationary eras. Demanding that the two periods of inflation solve the horizon problem leads to constraints on the number of e-folds which can be seen in Fig.~\ref{fig:HorizonConstraints} for two values of $\HI$ and $\HII$, respectively. The constraints resulting from solving the flatness problem are similar.

Let us quickly note two salient features. First, at very low scales of inflation, significantly less than the customary 60 e-folds are required. Second, a slightly ``too-short'' first period of inflation can be compensated for by a (relatively short) second inflationary period. However, for a significantly too-short first stage, the second period fully takes over and is the only observable one. In Fig.~\ref{fig:HorizonConstraints} this happens in the plateau regions on the left hand side of the plot.

We would like to stress that two-inflation is, in general, not degenerate with a late-inflation scenario. This is because the two-inflation scenario has a larger parameter space for the number of e-folds~$\NII$ since the horizon and curvature problems need not be solved by the second inflation alone. In a one-inflation scenario, the minimal number of e-folds follows directly from these constraints. This also completely fixes a possible dilution in the case when the axion field is already oscillating during inflation. On the other hand, the two-inflation scenario is more flexible because the flatness and horizon constraints can be addressed by a wide range of parameter combinations.

\subsection{Further details and validity}
The scale~$\mrm{a}{II}$ can be obtained by matching the temperature today at $a_o\equiv 1$ using the conservation of entropy with the effective entropic degrees of freedom $g_S(T)$~\cite[ch.~3.3]{Kolb:1990vq}:
\begin{align}
   \frac{\AII}{a_o} =  \left (\frac{g_S(\mrm{T}{CMB})}{g_S(\mrm{T}{II,reh})} \right )^{1/3} \left ( \frac{\mrm{T}{CMB}}{\mrm{T}{II,reh}} \right ) \approx \left (\frac{3.90}{g_S(\mrm{T}{II,reh})} \right )^{1/3} \left (  \frac{\SI{235}{\micro\electronvolt}}{\mrm{T}{II,reh}} \right ) \, .
\end{align}
Another important quantity is the temperature of the Universe right before and right after instant reheating. We will define these as $\mrm{T}{II}$ and $\mrm{T}{II,reh}$, respectively:
\begin{align}
   \mrm{\rho}{R}\left(\mrm{T}{II}\right) &= 3\mpl^2 \, \HII^2 \; \mathrm{e}^{-4\NII} \, , \label{eq:TII}\\
   \mrm{\rho}{R}\left(\mrm{T}{II,reh}\right) &= 3\mpl^2 \, \HII^2 \; \left ( 1 + \mathrm{e}^{-4\NII} \right ) \, , \label{eq:TrehII}
\end{align}
where the radiation energy density is $\mrm{\rho}{R}(T)=\uppi^2g(T)T^4/30$, with~$g$ being the effective number of degrees of freedom.

Our effective description of the background cosmology is, strictly speaking, only valid if the axion energy density does not dominate the background evolution of the Universe at any point before matter-radiation equality.

Let us identify when this assumption may be problematic. In the standard scenario, the Universe is radiation-dominated from the point of instantaneous reheating until matter-radiation equality. As QCD axions and ALPs have an (effective) equation of state $\langle w_A\rangle  < 1/3$, their energy density going back in time does not increase as fast as that of radiation. During the inflationary period, however, the energy density of the inflaton is constant, while the axion energy density increases for $a<\mrm{a}{II}$ as long as the field is dynamical. For $a<a_q$, the Universe becomes radiation-dominated again. If axions do not dominate the energy density at~$a_q$, there is no problem at earlier times.

For the phenomenologically interesting case of axions making up no more than $\Omega_{c,o}$, the only problematic time frame is $a_q<a<\mrm{a}{II}$. If QCD axions/ALPs are already in the oscillating regime where they behave like matter (cf., e.g., Sec.~\ref{sec:findings1}), their energy density will be suppressed compared to radiation by a factor of
\begin{equation}
	\frac{\rho_A}{\mrm{\rho}{R}} \lsim \frac{\mrm{\rho}{M}}{\mrm{\rho}{R}} \sim \frac{\mrm{T}{MR}}{T} \, .
\end{equation} 
Taking $T = \mrm{T}{II,reh}$ for the lowest $\HII$, we find a suppression factor of about $\num{7e-8}$. Turned around, if the energy density is suppressed less than this factor by the second stage of inflation, it should be still subdominant at~$a_q$ and no problems should arise. More generally, the smallest allowed suppression factor is approximately given by
\begin{equation}
\label{suppression}
	\num{7e-8} \, \sqrt{\frac{H_{\mathrm{II,\, min}}}{\HII}} \, \frac{\rho_{A,o}}{\rho_{c,o}} \, ,
\end{equation}
where we have allowed for the possibility that today's density in QCD axions/ALPs, $\rho_{A,o}$, is subdominant compared to today's total dark matter density, $\rho_{c,o}$.

For QCD axions this is automatically fulfilled for $\mrm{\theta}{i}\sim 1$ as long as $f_A \lsim \mathrm{few}\times\,\SI{e17}{\giga\electronvolt}$, according to Eq.~\eqref{eq:relicdensityapprox}. 

\begin{figure}
	\centering
	\includegraphics[width=0.618\linewidth]{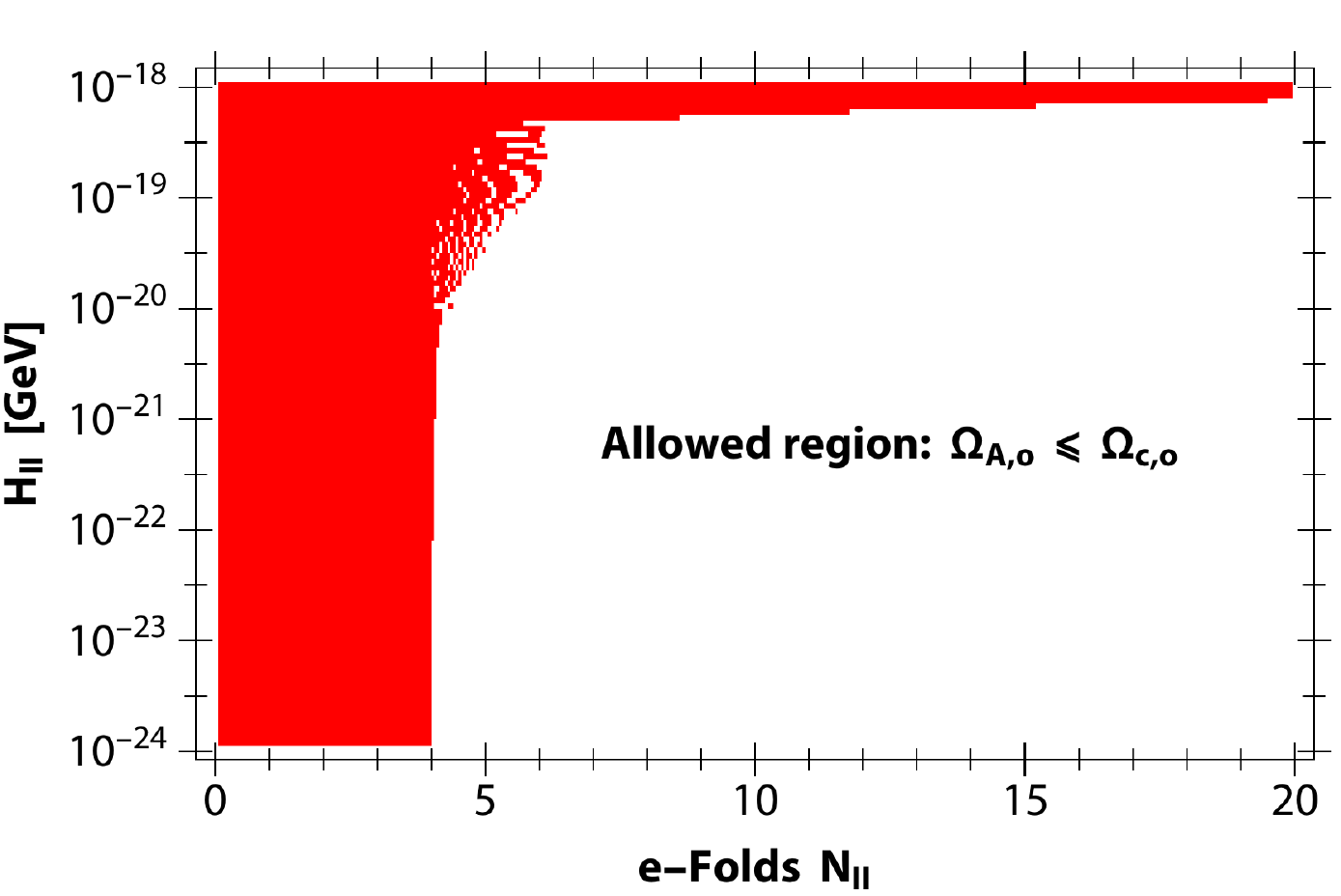}
	\caption{Density of axion DM. The red regions show where QCD axions with $f_A=\SI{e16}{\giga\electronvolt}$ and $\mrm{\theta}{i}=1$ have an abundance today of more than $\Omega_{c,o}h^2 \approx 0.12$~\cite{1502.01589}. The white region gives the viable parameter space. At the boundaries, we expect the axion to be all of DM. More details on the behaviour of the dilution by the second stage of inflation can be found in Sec.~\ref{sec:findings2}. \label{fig:qcdoverproduction}}
\end{figure}

In Fig.~\ref{fig:qcdoverproduction}, we show the regions in parameter space where, for $\mrm{\theta}{i} = 1$ and $f_A=\SI{e16}{\giga\electronvolt}$, the density is smaller than the observed DM density today (white regions). In these regions, two-inflation provides a good solution to the overproduction of DM for large values of $f_A$. We can see that this is possible in large regions of parameter space.

In the phenomenologically interesting region of Fig.~\ref{fig:qcdoverproduction} we can also estimate the contribution from QCD axions. For $f_A=\SI{e16}{\giga\electronvolt}$, we checked that QCD axions do not contribute more than about $0.1\%$ to energy density of the Universe during $a_q<a<\mrm{a}{II}$. For lower values of~$f_A$, the contribution is even smaller.

Indeed, our description in terms of scale factors and temperatures is largely independent of the equation of state. Therefore, most of our arguments still hold even if the QCD axion or ALP density is not always subdominant to the radiation and inflationary energy densities. Only the region around the time when field oscillations begin depends on the equation of state. If the energy density in QCD axion or ALPs is subdominant at these times, our results are at least qualitatively valid. This is a much weaker condition and nearly always fulfilled. 

It should also be noted that speaking of a second period of inflation may be misleading if the energy density during this period is ALP-dominated for some time such that the Universe is not necessarily undergoing accelerated expansion. For ALPs this is easily possible since their mass being independent of the scale $f_{A}$ allows for much larger initial densities.

A solution to the issue of axion energy density domination is to include $\rho_A$ as a contribution to~$H$. This would make the implementation more involved and less straight-forward to analyse the results.

\section{ALPs in the two-inflation scenario}\label{sec:findings1}
In this section and the next, we analyse the evolution of generic ALPs with a temperature-independent mass and QCD axions, respectively. In both cases, we have to solve the field equation~(\ref{eq:axionfieldapp}),
\begin{equation}
\ddot{\theta}+3H(T) \, \dot{\theta}+ m_A^2(T)\sin (\theta) = 0 \, ,\label{eq:axionfield}
\end{equation}
in order to obtain the energy density today, which is given by
\begin{equation}
\rho_{A,o}=\frac{1}{2} \, f_A^2 \, \left[\dot{\theta}^2_{o} + m_A^2(T_o) \, \theta^2_{o}\right] \, ,
\end{equation}
where at late times, indicated by the subscript ``$o$'', we can approximate the potential~(\ref{eq:axionpotential}) by its mass term. In the case of ALPs, it is quite straightforward to derive analytic expressions for the axion energy density today compared to the standard scenario. In the case of QCD axions, discussed in the next section, we employ numerical and semi-analytical methods in order to solve the field equation. More details on our procedure can be found in App.~\ref{app:WKB}. 

\subsection*{Dilution of the energy density for ALPs}
For the ALP case, we consider a temperature-independent, constant mass $m_A(T)\equiv m_{A,0}$. There are two qualitatively different regimes: If the Hubble scale of the second stage of inflation
is higher than the ALP mass, $3\HII\gsim m_{A,0}$, the field continues to be frozen during the entire second stage of inflation. After the second stage of inflation, however, the field evolution proceeds just as in the standard scenario with only one period of inflation. Therefore no change of density is expected.

The other more interesting case is $3\HII\lsim m_{A,0}$. Here, the field starts to oscillate between the two inflationary periods around a scale factor $\mrm{a}{osc,1}$ which is given by $3H(\mrm{a}{osc,1}) = m_{A,0}$. Using~(\ref{eq:Hcases}), we find that
\begin{equation}
   \frac{\mrm{a}{osc,1}}{\AII} = \left[\frac{m_{A,0}^2}{9\HII^2}-1\right]^{-1/4} \mathrm{e}^{-\NII} \, . \label{eq:ALPaosc}
\end{equation}
The expected relative dilution can be estimated by making a few approximations. In particular, let us assume that the axion field starts its oscillation when $3H\approx m_{A,0}$ and that the harmonic and adiabatic limits discussed in App.~\ref{app:WKB} directly apply. The latter holds as long as the initial $\mrm{\theta}{i}\ll 1$. For field values $\mrm{\theta}{i}\sim 1$ there are anharmonic effects to be taken into account for a more rigorous treatment which will introduce additional numerical factors as mentioned in~\cite{Turner:1985si, Lyth:1991ub, astro-ph/9405028, Bae:2008ue, 0903.4377}.
The value of this ratio is fixed at the scale factor $\mrm{a}{osc}^\mathrm{std}$ when the axion field starts to oscillate in the standard scenario. Given the scaling of the axion energy densities in the harmonic and adiabatic limits, the same initial energy densities, and using $m_{A,0}/(3\HII)\ll 1$, we find:
\begin{align}
   \frac{\Omega_{A,o}}{\Omega_{A,o}^\mathrm{std}} &= \frac{\rho_{A,o}^{}}{\rho_{A,o}^\mathrm{std}} \approx \left(\frac{\mrm{a}{osc,1}^{\phantom{\mathrm{std}}}}{\mrm{a}{osc}^\mathrm{std}}\right)^3 =
   \left(\frac{\mrm{a}{osc,1}}{\AII}\right)^3 \; \left(\frac{\AII}{\mrm{a}{osc}^\mathrm{std}}\right)^3\\
   &= \frac{g_S\(\mrm{T}{II,reh}\)g\(\mrm{T}{osc}^\mathrm{std}\)^{3/4}}{g_S\(\mrm{T}{osc}^\mathrm{std}\)g\(\mrm{T}{II,reh}\)^{3/4}} \left(\frac{m_{A,0}}{3\HII}\right)^{3/2} \left(1-\frac{m_{A,0}^2}{9\HII^2}\right)^{-3/4}\left(1+\mathrm{e}^{-4\NII}\right)^{-3/4} \; \mathrm{e}^{-3\NII}\\
   &\sim \mathrm{e}^{-3\NII} \, , \label{eq:ALPdilution}
\end{align}
where~(\ref{eq:TrehII}) has been used for the reheating temperature and we ignore the factors from the effective degrees of freedom in the last line. The result above agrees with numerical calculations. We also see that an additional phase of inflation dilutes the relative axion energy density as expected from earlier studies on late inflation~\cite{Dimopoulos:1988pw}.

\section{QCD axions in the two-inflation scenario}\label{sec:findings2}
The situation for QCD axions is more complicated than for ALPs. This is because the QCD axion mass has a strong temperature dependence due to its QCD origin. The form of that function has been parametrised in the literature using various techniques and can be approximated by a power law that turns into a constant around some critical temperature~$\mrm{T}{crit}$~\cite{Turner:1985si, Bae:2008ue, 0903.4377, 0910.1066},
\begin{equation}
   m_A(T) = m_{A,0} \,
   \begin{cases}
   \hfil 1 & \mathrm{for \; } T < \mrm{T}{crit} \\
   \left ( \mrm{T}{crit}/T \right )^{\beta/2} & \mathrm{for \; } T \geq \mrm{T}{crit} 
   \end{cases}\, . \label{eq:axionmass}
\end{equation}
To be explicit, we take $\beta = 6.68$, $m_{A,0} = \SI{61.1}{\micro \electronvolt} \left (\SI{E11}{\giga \electronvolt} / f_A \right )$ and $\mrm{T}{crit}=\SI{0.103}{\giga \electronvolt}$ from Wantz and Shellard~\cite{0910.1066}.\footnote{These values originate from interacting instanton liquid models~\cite{0908.0324,0910.1066}. Recent lattice calculations improve on this simple approximation~\cite{1606.03145, Borsanyi:2016ksw}.}

A consequence of~(\ref{eq:axionmass}) is the possibility of a drastic decrease of the axion mass at reheating. In general, this leads to two distinct cases. If the reheating temperature is small and the mass does not (or only slightly) decrease, the field continues its oscillations~(\emph{Case~1}). In this case we observe the expected reduction in energy density similar to the case of ALPs. For higher reheating temperatures however, the axion becomes very light~(\emph{Case~2}) and oscillations stop. Depending on the time derivative of the axion field when this happens, the field can either be directly frozen or it continues to move with a velocity damped by the Hubble drag. More generally, the field evolution will depend strongly on the phase of the oscillation of the axion field at which reheating sets in. 
As we discuss below both a significant dilution as well as a small increase in density is possible.

In this more complicated situation, we solve the field equations numerically using the procedure described in App.~\ref{app:WKB}.

\subsection{Characterising regimes of different behaviour}
The different regimes can be characterised by the ratio of the temperature-dependent mass to the Hubble scale,
\begin{align}
   \mu(T) \equiv \frac{m_A(T)}{3H(T)} = m_A(T) \begin{cases}
   \[9\mrm{H}{II}^2 + 3\mrm{\rho}{R}(T)/\mpl^2\]^{-1/2} & \text{before reheating}\\
   \[3\mrm{\rho}{R}(T)/\mpl^2\]^{-1/2} & \text{after reheating}
   \end{cases} \, , \label{eq:muT} 
\end{align}
where $m_A(T)$ is given by~(\ref{eq:axionmass}). This quantity can be used to discriminate different cases by inserting the relevant temperatures.

\begin{figure}
	\centering
	\includegraphics[width=0.85\linewidth]{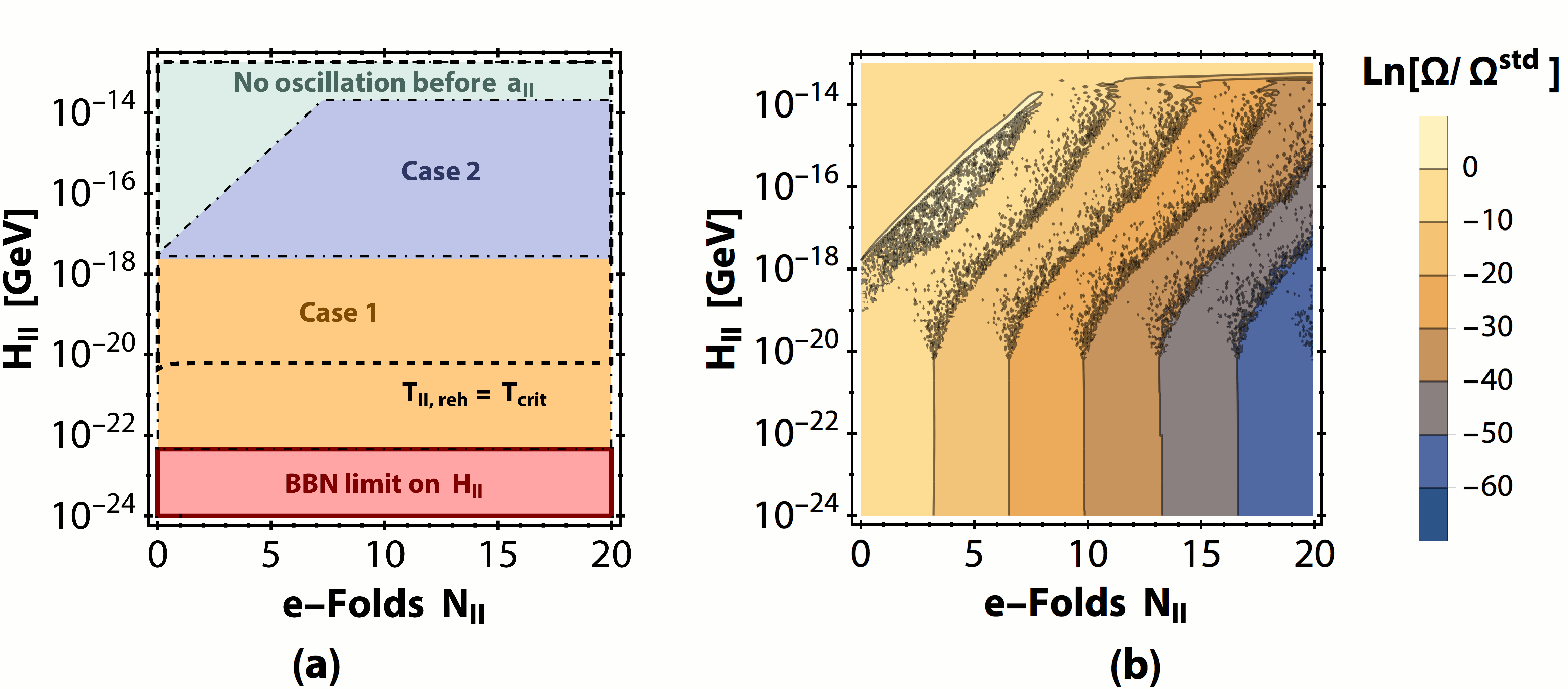}
	\caption{Results for the QCD axion energy density compared to standard cosmology. Subfigure~(a) shows the different cases that can occur for this scenario as explained in the main text. Plot~(b) shows -- in order to compare to (\ref{eq:ALPdilution}) -- the natural logarithm of the ratio between the energy densities today in the two-inflation and in the one-inflation scenario, for $f_A=\SI{E11}{\giga \electronvolt}$ and $\mrm{\theta}{i}=\num{e-5}$.\label{fig:QCDaxionresults}}
\end{figure}

A non-trivial modification of the standard scenario requires that the field starts to oscillate before the end of the second inflation. Using definition~(\ref{eq:muT}), the condition $\mu(T)=1$ must be fulfilled before~$\AII$. This is guaranteed if $\mu\left(\mrm{T}{II}\right) < 1$, where we remember that $\mrm{T}{II}$ at the end of inflation but just before reheating. 
For our benchmark value of $f_{A}=\SI{E11}{\giga\electronvolt}$, the relevant region is shown in Fig.~\ref{fig:QCDaxionresults}(a)
below the green area. The dependence on the number of e-folds in Fig.~\ref{fig:QCDaxionresults}(a) arises because the temperature continues to drop during inflation. As the temperature-dependent axion mass increases accordingly, longer periods of inflation permit oscillations for larger values of $\HII$.

Indeed, the quantitative amount of dilution will depend on the relation between the temperature when oscillations start during inflation, $\mrm{T}{osc,1}$, and the temperature at the start of the second period of inflation, $T_q$. If the axion field oscillations begin during the second inflationary period, the relative dilution of axion energy density is reduced because it does not happen for the full duration of inflation.

A second qualitatively important factor is whether the reheating temperature is sufficiently high such that the axion mass essentially vanishes and the oscillations stop, $\mu\left(\mrm{T}{II,reh}\right) = 1$ (e.g.~$\mrm{T}{II,reh}=\SI{1.44}{\giga\electronvolt}$ in case of $f_A=\SI{e11}{\giga\electronvolt}$). This delineates the boundary between our \emph{Case 1} and \emph{Case 2} as already mentioned above. Numerical values for two examples can be found in Table~\ref{tab:regions}.
\begin{table}
	\caption{Numerical values for $\HII$ (in the limit of $\NII\gg 1$) to determine the boundaries of various regions of interest in parameter space for a large number of e-folds. The dependence on the axion decay constant arises because the axion mass depends on it.\vspace{0.5em}\label{tab:regions}}
	\centering
	\begin{tabular}{@{}llrr@{}}\toprule
		Description & Equation & \multicolumn{2}{c}{$\log_{10}\left(\HII/\si{\giga\electronvolt}\right)$}\\ \cmidrule{3-4}
		& & $f_A=\SI{e11}{\giga\electronvolt}$ & $f_A=\SI{e16}{\giga\electronvolt}$\\ \midrule
		No oscillation before $\AII$ & $\mu\left(\mrm{T}{II}\right) < 1$ & $>-13.7$ & $>-18.7$\\
		\emph{Case 1} & $\mu\left(\mrm{T}{II,reh}\right) \gsim 1$ & $\lsim -17.6$ & $\lsim -19.5$\\
		\emph{Case 2} & $\mu\left(\mrm{T}{II,reh}\right) \lsim 1$ & $\gsim -17.6$ & $\gsim -19.5$ \\
		\bottomrule
	\end{tabular}
\end{table}
Finally, the red region in Fig.~\ref{fig:QCDaxionresults}(a) shows the lower limit on $\HII$ from requiring that the second inflation does not interfere with BBN.

\subsection{Quantitative changes in the density}
We are now ready to compare the actual dilution in the different regimes. In particular, we focus on the case of sufficiently small initial field values such that $|\theta|\ll \otherpi$ at all times and the potential can be taken as approximately quadratic. In this regime, the resulting density ratios to the standard scenario are independent of~$\mrm{\theta}{i}$. For our choice of $\mrm{\theta}{i}=\num{e-5}$, Equation~(\ref{eq:relicdensityapprox}) shows that any sub-Planckian value of~$f_A$ is allowed. However, this choice is mostly to simplify the discussion and to stay in the regime of $|\theta| < \otherpi$ at all times of the evolution, given our benchmark value of $f_{A}=\SI{E11}{\giga\electronvolt}$. We comment on the case of larger field values, where $|\theta|>\otherpi$, in Sec.~\ref{largefields}.

Fig.~\ref{fig:QCDaxionresults}(b), shows the numerically obtained axion density today compared to the standard scenario depending on the Hubble scale of our second inflation and the number of e-folds.
Let us now understand the salient features in more detail.

\subsubsection*{Case 1. Low reheating temperature: oscillations during second inflation and reheating}
For \emph{Case~1}, i.e. in the lower region of Fig.~\ref{fig:QCDaxionresults}, below the dashed line $\mrm{T}{II, reh}=\mrm{T}{crit}$ in Fig.~\ref{fig:QCDaxionresults}(a), the relative axion energy density is diluted by about a factor of $\mathrm{e}^{-3\NII}$ as it is the case for ALPs. This is as expected since the field continues to oscillate with a constant mass and therefore behaves like matter during the whole second stages of inflation and reheating. 

Let us now consider higher values of the Hubble scale and consequently also to higher values of the reheating temperature and, in particular, values of the reheating temperature lying above the dashed line $\mrm{T}{II, reh}=\mrm{T}{crit}$ in Fig.~\ref{fig:QCDaxionresults}(a). While per definition of our \emph{Case~1} the field continues to oscillate, the mass nevertheless decreases at reheating. As we will discuss in more detail below for the case where the oscillations stop, the effect on the density now strongly depends on the oscillation phase at which reheating sets in. 
This is why Fig.~\ref{fig:QCDaxionresults}(b) exhibits a somewhat ``noisy'' behaviour. To get some understanding let us quickly look at two special cases. If the field value is at a turning point of the oscillation, the amplitude remains unchanged. However, if the field is at zero and all the energy is kinetic energy, the amplitude of the field when oscillations resume after reheating, $|\theta|_\mathrm{max}$, can be enhanced to
\begin{equation}
\left|\theta\right|_\mathrm{max} \sim \frac{\left|\mrm{\dot{\theta}}{II,max}\right|}{m_{A}(T_\mathrm{II,reh})}\sim \frac{m_{A}(T_\mathrm{II,reh})}{m_{A}(T_\mathrm{II})}\left|\theta_\mathrm{II,max}\right| \, ,
\end{equation}
where $|\theta_\mathrm{II,max}|$ indicates the amplitude of the oscillating field just before the end of the inflationary period.

\subsubsection*{Case 2. High reheating temperatures: Oscillations stop at reheating and resume only later}
Considering a fixed number of e-folds, the dilution is reduced when we go to higher values of the Hubble scale~$\HII$. This qualitative trend can be seen from the right-hand panel of Fig.~\ref{fig:QCDaxionresults} and can be understood quite easily: At small values of the Hubble scale, i.e. \emph{Case~1} discussed above, the axion field oscillates during the whole inflationary period. Its density is reduced by the full expansion factor. At higher Hubble scale, the temperature at the start of the second inflation is not yet below the critical temperature and the temperature-dependent axion mass can be below the Hubble scale. The field is frozen and no dilution occurs until the point when the temperature has dropped sufficiently. The field oscillates and is diluted only through part of the inflationary period. Finally, when the Hubble scale becomes bigger than the vacuum mass, no oscillation and no dilution happens before the end of the second stage of inflation.

\begin{figure}
	\centering
	\includegraphics[width=0.618\linewidth]{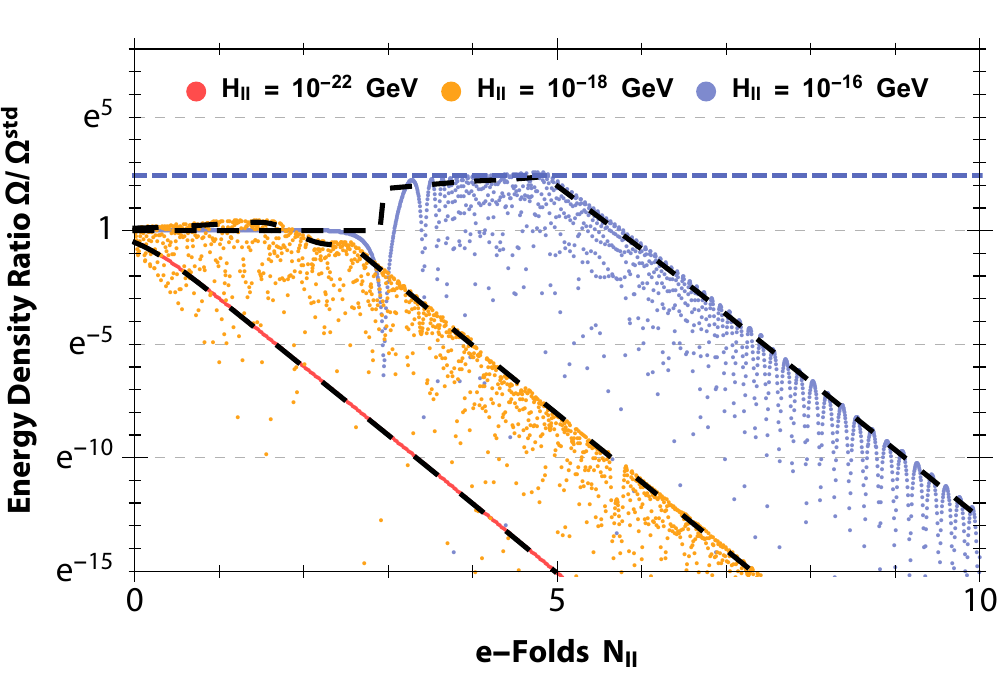}
	\caption{Iso-$\HII$ lines for Fig.~\ref{fig:QCDaxionresults}(b) to illustrate the different cases and show the ``scanning effect'' of $\NII$ ($f_A=\SI{E11}{\giga \electronvolt}$, $\mrm{\theta}{i}=\num{e-5}$). The black dashed lines indicate the envelope estimated from the maximal value for the relative energy density in \emph{Cases~1} and~\emph{2}. The blue dashed line is the maximal possible value in \emph{Case~2} from~(\ref{eq:maxincrease}).\label{fig:QCDaxionttrace}}
\end{figure}

Beyond this general trend, the figure also exhibits a noisy behaviour similar to the one observed above at the upper end of \emph{Case~1}. To resolve this, we plot in 
Fig.~\ref{fig:QCDaxionttrace} the change in density compared to the standard scenario as a function of the number of e-folds for several values of the Hubble scale. For the smallest value (red curve), we find the dilution expected from~(\ref{eq:ALPdilution}). For higher values (yellow and blue curves), we observe the phase-induced noisiness and even an oscillating behaviour. Indeed, even a moderate increase in the energy density ratio is possible.
Let us now understand these observations in more detail.

We first point out that the most interesting region is where the reheating temperature is big enough such that the axion can be viewed as effectively massless just after inflation.
Using this approximation, the field values develop according to~(\ref{eq:axionasymsolrad}),
\begin{equation}
\theta (a) = \mrm{\theta}{II} + \frac{\mrm{\dot{\theta}}{II}}{\mrm{H}{II}\sqrt{1 + \mathrm{e}^{-4\NII}}}\left ( 1 - \frac{\mrm{a}{II}}{a} \right )\, , \label{eq:axionsol2}
\end{equation}
where $\mrm{\theta}{II}\equiv \theta (\AII)$ and $\mrm{\dot{\theta}}{II}\equiv\dot{\theta}(\AII)$ are the field value and its time derivative at the very end of inflation. 

Let us now again consider the two special cases where the field at the end of inflation is either at the maximum of the oscillation or at zero. The first case corresponds to an initially vanishing time derivative, the second to a vanishing initial field value.

In the first case, the field is effectively frozen until it starts to oscillate again when the temperature has dropped sufficiently. The corresponding density is that of a standard one-inflation scenario with an initial field value given by the field value at the end of the second stage of inflation, $\mrm{\theta}{II}\equiv \theta (\AII)$.
Depending on the length of the oscillating phase during the second inflation, this field value is reduced compared to the initial field value at the end of the first stage of inflation.

\begin{figure}
	\centering
	\includegraphics[width=0.618\linewidth]{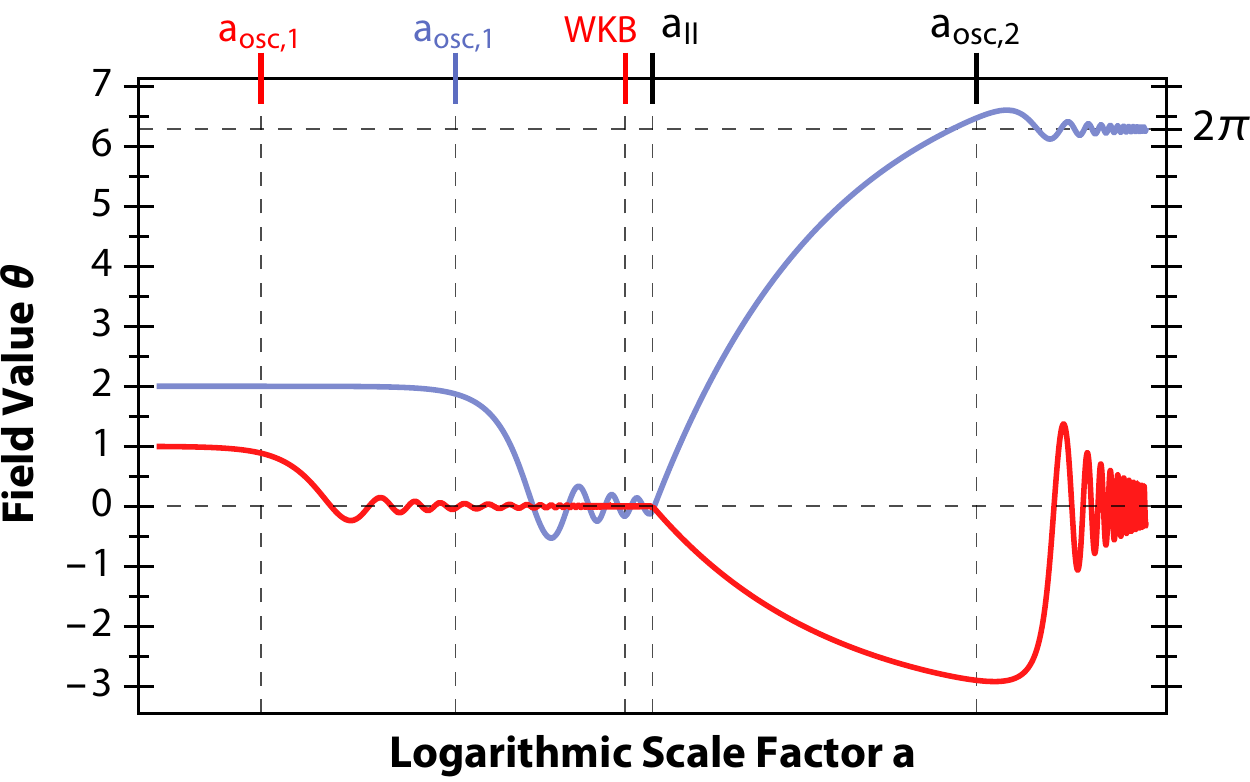}
	\caption{Example for the evolution of an axion field with $f_A=\SI{e11}{\giga\electronvolt}$ and $\HII=\SI{e-16}{\giga\electronvolt}$, while $\NII=4$ (blue curve) or $\NII=5$ (red curve). The label ``WKB'' denotes the point where the semi-analytical WKB-inspired ansatz takes over from the numerical solution. Note that for the chosen values of $\mrm{N}{II}$, the oscillation re-commence essentially ar the same scale factor after reheating. \label{fig:qcdaxionoscex}}
\end{figure}

The second case is more interesting. Here, the field evolution according to~\eqref{eq:axionsol2} can be described as an asymptotic growth of the absolute axion field value~$\left|\theta\right|$ because of the scale factor dependence in the second term in~\eqref{eq:axionsol2}. This is illustrated by two examples in Fig.~\ref{fig:qcdaxionoscex}.
As we can see from the figure, the growth in the field value compared to the typical amplitude at the end of the second stage of inflation can be quite significant.
We can estimate that\footnote{If the field before the second stage of inflation is not too small it is actually possible that this behaviour can increase a very small $\mrm{\theta}{II}$ to values $\theta\gsim\otherpi$ as seen in Fig.~\ref{fig:qcdaxionoscex}. Mathematically speaking, the periodicity of the axion potential defines equivalency classes on the field values, which should be represented by, e.g., the equivalent value in the range $(-\otherpi,\otherpi]$. For a clearer presentation, we ignore this subtlety and do not map values outside of this range back to their equivalent values.}
\begin{equation}
\left|\theta\right|_\mathrm{max} \sim \frac{\left|\mrm{\dot{\theta}}{II,max}\right|}{\HII}\sim \frac{m_{A}(T_\mathrm{II})}{\HII}\left|\theta_\mathrm{II,max}\right|,
\end{equation}
where, as before, $|\theta_\mathrm{II,max}|$ indicates the amplitude of the oscillating field just before the end of the inflationary period. Since the mass can be significantly bigger than the Hubble scale this can be a sizeable enhancement, as is also visible in Fig.~\ref{fig:qcdaxionoscex}.
We note, however, that this enhancement does not directly translate into a corresponding enhancement compared to the standard single inflation scenario. The reason is that the field must already be oscillating at the end of the second inflation and is therefore also affected by a corresponding dilution during the inflation.
To some degree, the dilution during the second stage of inflation and the enhancement after reheating compensate each other.

So far we have considered two limiting cases. In general, the behaviour determined by~\eqref{eq:axionsol2}
depends on the phase of the axion field oscillation at the time of reheating. In our simple model this is fixed by the number of oscillations before the reheating sets in. For a given Hubble scale, this depends on the number of e-folds. This explains the ``scanning'' effect with the number of e-folds $\NII$ and the noisiness of Fig.~\ref{fig:QCDaxionresults}(b).

Before we turn to an estimate of the maximal possible enhancement, we note that -- while the case of the field starting with maximal initial velocity essentially gives the case of maximal enhancement -- the case of maximal field value does not correspond to the situation with maximal dilution. More dilution can be achieved if the initial conditions in~\eqref{eq:axionsol2} are chosen such that the field is close to zero at the time when oscillations resume.

\subsubsection*{Deviations from dilution: Maximal possible increase}
Let us obtain an estimate for the maximal possible enhancement of today's energy density in axions compared to the standard scenario. The field first starts oscillating at a temperature~$\mrm{T}{osc,1}$ before the end of the second inflation. Afterwards, the Universe reheats and the axion mass decreases. As already explained above, we expect maximal enhancement if $\mrm{\theta}{II}\approx 0$ such that reheating occurs when the axion field has maximal velocity,
\begin{equation}
	\left|\mrm{\dot{\theta}}{II}\right| \approx m_{A}(T_\mathrm{II})\left|\theta_\mathrm{II,max}\right| \, , \label{eq:ThetaDotMaxKinErg}
\end{equation}
with field amplitude $\theta_\mathrm{II,max}$ just before reheating. Assuming adiabatic evolution we can connect the above equation with the initial misalignment angle~$\mrm{\theta}{i}\ll 1$ via~(\ref{eq:edescaling}):
\begin{equation}
\mrm{\dot{\theta}}{II}^2 \approx m_A\(\mrm{T}{osc,1}\) \, m_A\(\mrm{T}{II}\) \, \left(\frac{\mrm{a}{osc,1}}{\AII}\right)^3 \, \mrm{\theta}{i}^2 \, . \label{eq:ergscaling}
\end{equation}
From~(\ref{eq:axionsol2}), we find for the evolution of~$\theta$ after reheating that
\begin{equation}
\theta (a) \approx \frac{\mrm{\dot{\theta}}{II}}{\HII} \left(1 - \frac{\AII}{a}\right) \, . \label{eq:SolThetaAfter}
\end{equation}
In order to find the maximal enhancement, we have to estimate the value the field has reached at the time and corresponding temperature, $T_\mathrm{osc, 2}$, when the field starts oscillating again. From this point on, the evolution is exactly the same as in the one-inflation scenario. We can therefore compare the densities/field values directly at this point.
Combining~\eqref{eq:SolThetaAfter} and~\eqref{eq:ergscaling}, we obtain,
\begin{equation}
	\theta^2(T_\mathrm{osc, 2}) \sim \frac{\mrm{\dot{\theta}}{II}^2}{\HII^2} \left(1 - \frac{\AII}{\mrm{a}{osc,2}}\right)^2 \approx \frac{m_A\(\mrm{T}{osc,1}\) \, m_A\(\mrm{T}{II}\)}{\HII^2} \, \left(\frac{\mrm{a}{osc,1}}{\AII}\right)^3 \left(1 - \frac{\AII}{\mrm{a}{osc,2}}\right)^2\, \mrm{\theta}{i}^2 \, . \label{eq:maxaxwandering}
\end{equation}
In Fig.~\ref{fig:QCDaxionttrace}, we show the change in the energy density as a black, dashed line for every value of $\NII$ where \emph{Case~2} applies and also show the result that one obtains for \emph{Case~1} in a similar fashion. Indeed, the results match the upper envelope for all curves.
 
We can also obtain an upper limit for the maximal enhancement for a given value of $\HII $. 
To do this we have to consider the two possible situations depending on whether $\mrm{T}{II}$ is greater or smaller than $\mrm{T}{crit}$. In the latter case we have $m_A\(\mrm{T}{II}\)=m_{A,0}$. This corresponds to the drop-off behaviour at large values of $\NII$ that can be seen in Fig.~\ref{fig:QCDaxionttrace}. Let us now consider the situation $\mrm{T}{II}\gsim \mrm{T}{crit}$. In this case we have to use the temperature-dependent mass~(\ref{eq:axionmass}). Combining $m_A\(\mrm{T}{osc,1}\)\sim 3\HII$ with~(\ref{eq:maxaxwandering}), we find
\begin{equation}
	\(\frac{\Delta\theta}{\mrm{\theta}{i}}\)^2 \lsim 9 \, \frac{m_A\(\mrm{T}{II}\)}{m_A\(\mrm{T}{osc,1}\)} \, \left(\frac{\mrm{a}{osc,1}}{\AII}\right)^3 \left(1 - \frac{\AII}{\mrm{a}{osc,2}}\right)^2 \sim 9 \, \left(\frac{\mrm{a}{osc,1}}{\AII}\right)^{3-\beta/2} \left(1 - \frac{\AII}{\mrm{a}{osc,2}}\right)^2
\end{equation}
The maximum value of this enhancement occurs when $\mrm{T}{II}\approx \mrm{T}{crit}$. Using the appropriate temperature dependence, $m_{A,0}=m_A\(\mrm{T}{osc,1}\)\sim 3\HII$ we have for the maximum enhancement at this point,
\begin{equation}
	\left. \frac{\Omega_{A,o}}{\Omega_{A,o}^\mathrm{std}}\right|_\mathrm{max} \sim \; 9 \, \left(\frac{m_{A,0}}{3\HII}\right)^{0.17}\left(1-\left(\frac{\mrm{H}{II, \, min}}{\HII}\right)^{1/2}\right)^2 \, , \label{eq:maxincrease}
\end{equation}
where, $H_\mathrm{II,\, min}$ denotes the minimal value for \emph{Case~2} to occur. The result for $\HII = \SI{e-16}{\giga\electronvolt}$ is about $\mathrm{e}^{2.44}$ and has been included in Fig.~\ref{fig:QCDaxionttrace} as a blue, dashed line. This demonstrates that the increase in relative energy density is relatively limited.

\subsection{Towards more realistic reheating scenarios}\label{realistic}
\begin{figure}
	\centering
	{
	\includegraphics[width=0.6\linewidth]{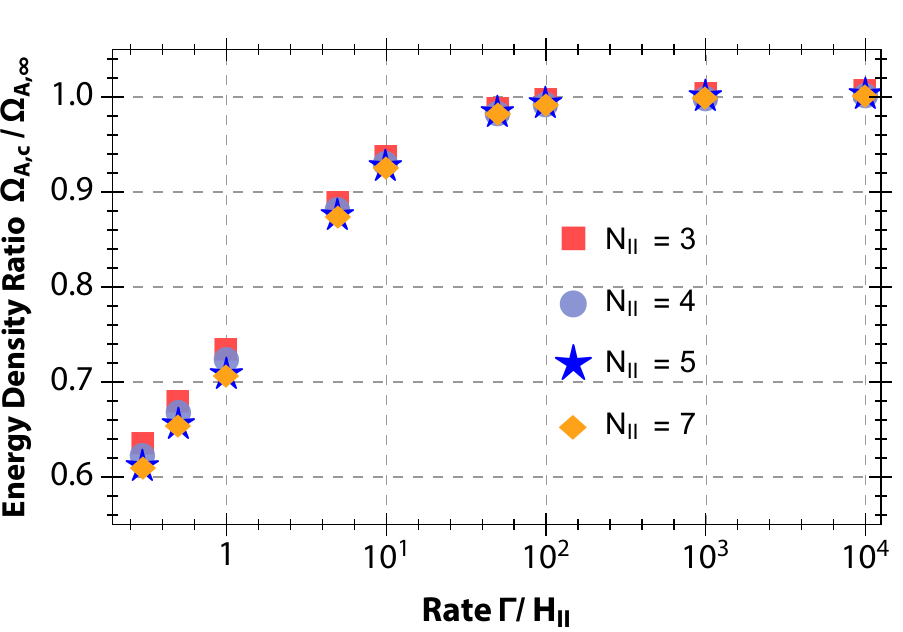}
	}
	\caption{Dependence of the axion energy density today on the decay constant $c\equiv\Gamma/\HII$ of the inflaton. The parameters are $f_A=\SI{e11}{\giga\electronvolt}$, $\HII=\SI{e-16}{\giga\electronvolt}$ and $\mrm{\hat{\theta}}{II}=0.002$.
	\label{fig:cdependence}}
\end{figure}

One might wonder if the observed behaviour of the axion field in \emph{Case~2}, and in particular the observed potential for an increase in the density, is just an artefact of instant reheating. 
In order to investigate this, we change our model such that reheating sets in at $t_\mathrm{II}$ with a finite energy transfer rate $\Gamma$ into relativistic degrees of freedom (cf. e.g.~\cite[ch.~4.2]{Weinberg:2008zzc}),
\begin{align}
\mrm{\rho}{Inf} &= \mrm{\rho}{II} \, \(\frac{\AII}{a(t)}\)^3 \,  \mathrm{e}^{-\Gamma \(t-\tII\)} \, ,\\
H &= \sqrt{\frac{\mrm{\rho}{Inf}+\mrm{\rho}{R}}{3\mpl^2}} = \frac{\dot{a}}{a} \, ,\\
\mrm{\dot{\rho}}{R} &= -4H\mrm{\rho}{R} + \Gamma \mrm{\rho}{Inf} \, .
\end{align}
For simplicity we just consider the case where the field velocity at the beginning of reheating is maximal,\footnote{As discussed above, the phase is strictly speaking fixed by the value of $\NII$. To make the results comparable, we have imposed by hand that the field velocity is maximal.} corresponding to maximal enhancement in the instantaneous case. Fig.~\ref{fig:cdependence} shows examples for the resulting behaviour compared to the result of instantaneous reheating.

As expected, for $\Gamma \gg \HII$, the instant reheating behaviour is recovered. 
For finite reheating rate the resulting density is slightly smaller, decreasing with decreasing rate. This can be understood as follows.
In the example of Fig.~\ref{fig:cdependence} we have $\mrm{T}{II}<\mrm{T}{crit}$ for $\NII\gsim 4.8$ and, for instant reheating, the temperature would increase to about $\SI{8.9}{\giga\electronvolt}$. This means that even in the case of not very efficient reheating, the axion can become very light very soon after the second period of inflation.
Indeed, one can check that even for moderate reheating rates of order $\HII$, the reheating is so fast that the field turns essentially massless in less than an oscillation period.\footnote{One should remember that $\rho_\mathrm{R}\sim T^4$ and therefore the temperature rises very quickly at the beginning of reheating.} The dominant effect reducing the density is actually the slightly increased Hubble friction. The latter comes about because of the slower decrease in the total energy density at the beginning of reheating as only a part of it is made of radiation, which is subject to dilution.

\subsection{Larger field values}\label{largefields}
So far we have only considered very small initial field values and indeed required that the field value remains sufficiently small at all times during the evolution. Yet, natural values for the initial misalignment angle are $\mrm{\theta}{i}\sim 1$. In fact, if the PQ phase transition occurs after the first stage of inflation, values or order unity are unavoidable. Let us therefore briefly comment on this situation while leaving a more detailed analysis to future work.
\begin{figure}
	\centering
	{
		\includegraphics[width=0.85\linewidth]{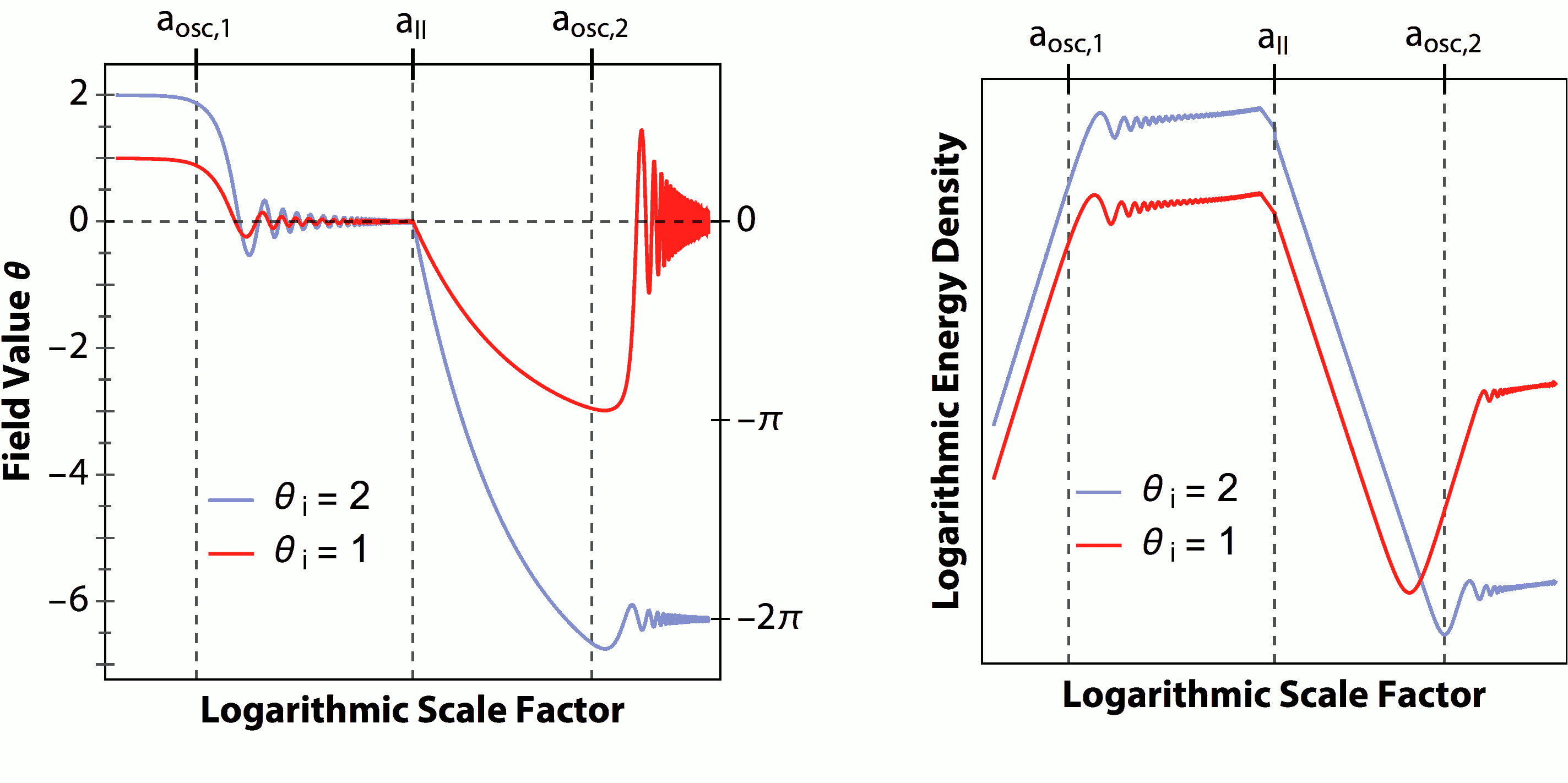}
	}
	\caption{An example for how leaving the field range from $-\otherpi$ to $\otherpi$~(left) can result in a lower energy density today~(right).  The energy density for $\mrm{\theta}{i}=2$ is initially about a factor~3 higher than for $\mrm{\theta}{i}=1$ (anharmonic effects). After the second oscillations, this ratio changes to about~0.02. The other parameters in this example are $f_A=\SI{E11}{\giga\electronvolt}$, $\HII=\SI{E-16}{\giga\electronvolt}$ and $\NII = 5$. \label{fig:largefield}}
\end{figure}
There are two main effects of larger field values. The first is the change in the evolution from the anharmonicity of the potential. This effect also exists in a standard single inflation scenario. The second is more interesting. We have already found that field values can increase if the reheating temperature is large enough. For sizeable $\mrm{\theta}{i}$, the field values can now leave the range $(-\otherpi,\otherpi]$ and start to explore the periodicity of the axion potential (cf. Fig.~\ref{fig:qcdaxionoscex}). A larger initial field value now may actually lead to a reduced density. An example of this behaviour is depicted in Fig.~\ref{fig:largefield}, where the amplitudes of field oscillations after reheating are bigger for the field with smaller~$\mrm{\theta}{i}$.
\begin{figure}
	\centering
	{
	\includegraphics[width=0.618\linewidth]{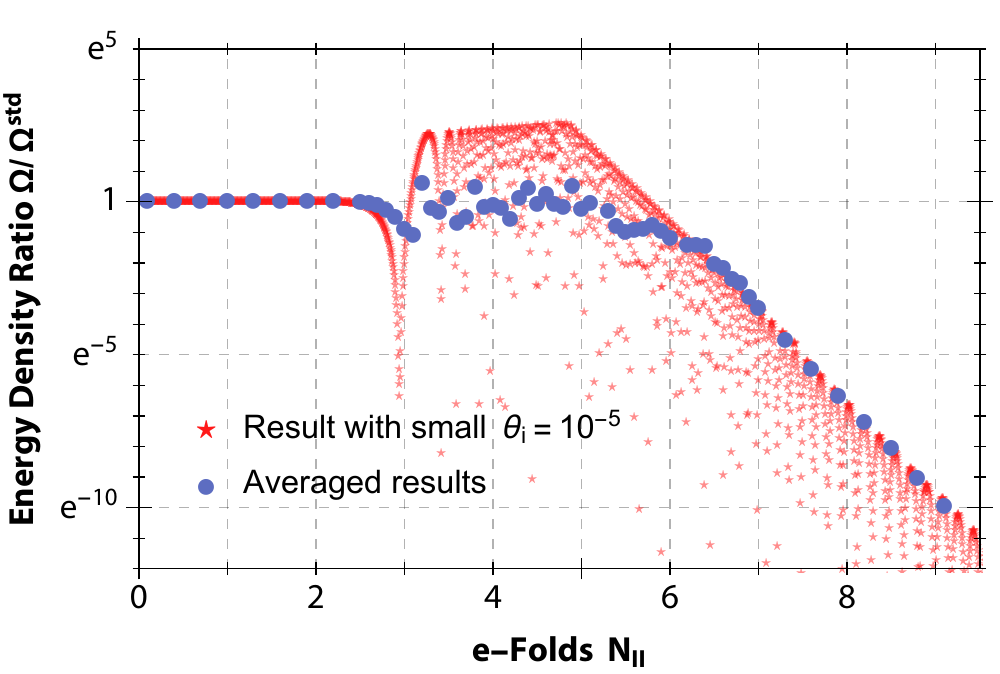}
	}
	\caption{Comparison of the small-angle approximation with $\mrm{\theta}{i}=\num{e-5}$ and post-inflationary PQ breaking. For the latter, we show the average energy density from 1000~samples in the interval $(-\otherpi,\otherpi]$. The parameter values are $f_A=\SI{e11}{\giga\electronvolt}$ and $\HII=\SI{e-16}{\giga\electronvolt}$.\label{fig:average}}
\end{figure}

An interesting question is also when the field has different values for the misalignment angle in different regions of the Universe, as one would expect in a situation where the PQ transition occurs after the initial stage of inflation. In this case a first estimate can be obtained by averaging over different initial values of $\theta$. In Fig.~\ref{fig:average}, we show the ratio between this average density in the standard and in the two-inflation scenario. To guide the eye, we also show for the same values of parameters the corresponding ratio in the case where the initial condition is a homogeneous small value.

\section{Conclusions}\label{sec:discussion}
We have revisited the possibility of a second episode of inflation as a potential way to avoid overproduction of QCD axions and ALPs via the misalignment mechanism. In agreement with previous work~\cite{Dimopoulos:1988pw, 1507.08660}, we find for ALPs with constant mass that the two-inflation scenario generally dilutes the energy density if the field is already oscillating during the second inflationary period. 
However, if the mass is temperature-dependent, as for the QCD axion, the behaviour can be more complex than a simple dilution effect. 
In fact, the relative axion energy density compared to the standard scenario may even be increased by a moderate amount. This somewhat surprising result can be explained by the fact that if the reheating temperature is above the QCD phase transition temperature, the axion mass drops significantly at reheating. If this happens when the field has maximal velocity, it can reach a much larger value before it is stopped either by the diminished potential or Hubble friction. While the increase compared to a one-inflation scenario is moderate, it is a rather large deviation from the na{\"i}vely expected dilution effect in a two-inflation scenario.

In this paper, we have mainly focused on situations where the initial field value is rather small. However, one of the most attractive scenarios for axion dark matter is that the PQ phase transition occurs after inflation (or in our case after the first stage of inflation). In this case, the currently observable universe consists of a huge number of initially causally disconnected regions, where the field values are randomly chosen from the range $-\otherpi$ to $\otherpi$. The dark matter density from the misalignment mechanism is then usually obtained as an average over these randomly chosen values. Importantly, the typical field values are then usually not small.
As briefly discussed in Sec.~\ref{largefields}, this can lead to interesting behaviour. In such a scenario, one also expects large inhomogeneities which are potentially observable as so-called miniclusters~\cite{Hogan:1988mp,Kolb:1993zz,Kolb:1993hw,Fairbairn:2017sil,Enander:2017ogx}. These too could possibly be modified by the second stage of inflation. In particular, they may be blown up, potentially changing observable signatures such as gravitational lensing effects~\cite{Fairbairn:2017dmf,Fairbairn:2017sil}. In this way, one may have an opportunity to directly probe the inflationary scenario.
The same applies to topological defects such as axion strings and domain walls, which can contribute significantly to the dark matter density via their decay. Their contribution may also behave non-trivially in a two-inflation scenario. The case of larger field values therefore presents an interesting area for future investigations.

\section*{Acknowledgements}\label{sec:acknowledgements}
JJ would like to thank J.~Redondo for as usual helpful and joyful discussions. This paper is partly based on the Master's thesis of SH~\cite{HoofMsc2015}, who also gratefully acknowledges funding by the Imperial College President's PhD Scholarship scheme.
JJ is supported by DFG TR33 ``The Dark Universe'' as well as the European Union Horizon 2020 research and innovation under the Marie Sk{\l}odowska-Curie grant agreement Numbers 674896 and 690575.

\appendix

\section{Numerical and analytical solution of the axion field equation}\label{app:WKB}
The equation for a homogeneous scalar field~$A(t) \equiv f_A \, \theta(t)$ in a universe with Hubble parameter $H$ is given by~\cite[app.~B]{Weinberg:2008zzc}:
\begin{equation}
\ddot{\theta} + 3H\dot{\theta} + \frac{V'[\theta]}{f_A^2} = 0 \, , \label{eq:scalarfield}
\end{equation}
 where $V$ is the periodic axion potential, canonically written as
\begin{equation}
V[\theta] = f_A^2m_A^2(T) \left[1 - \cos (\theta)\right] \, , \label{eq:axionpotential}
\end{equation}
with temperature-dependent axion mass~$m_A(T)$. Combing equation~(\ref{eq:scalarfield}) with the potential in~(\ref{eq:axionpotential}) yields:
\begin{equation}
\ddot{\theta}+3H(t) \, \dot{\theta}+ m_A^2(T)\sin (\theta) = 0 \, . \label{eq:axionfieldapp}
\end{equation}
If Hubble damping is much larger than the mass term, $m_A \ll 3H$, one can neglect the latter in~(\ref{eq:axionfieldapp}) and only the following equation needs to be solved:
\begin{equation}
\ddot{\theta} + 3H\dot{\theta} = \ddot{\theta} + \frac{3\dot{a}}{a} \, \dot{\theta} \approx 0 \, .
\end{equation}
This can be integrated with initial conditions $\theta(\mrm{t}{i})\equiv \mrm{\theta}{i}$, $\dot{\theta}(\mrm{t}{i})\equiv \mrm{\dot{\theta}}{i}$, and defining $a(\mrm{t}{i}) \equiv \mrm{a}{i}$:
\begin{equation}
\theta(t) = \mrm{\theta}{i}^{\phantom{3}} + \mrm{\dot{\theta}}{i}^{\phantom{3}} \! \mrm{a}{i}^3 \int_{\mrm{t}{i}}^{t}\! a^{-3}(\tau) \, \mathrm{d}\tau = \mrm{\theta}{i}^{\phantom{3}} + \mrm{\dot{\theta}}{i}^{\phantom{3}} \! \mrm{a}{i}^3 \int_{\mrm{a}{i}}^{a}\! \frac{1}{H(\alpha)\alpha^4} \, \mathrm{d}\alpha \, . \label{eq:generalsol}
\end{equation}
The second term in the above expression is often rejected because it is singular for $a\rightarrow 0$~\cite[p.~198]{Weinberg:2008zzc} or by assuming that $\mrm{\dot{\theta}}{i}/\mrm{H}{i} \ll 1$~\cite[p.~429]{Kolb:1990vq}. This may be justified for early times and the first oscillations of the axion field but, in general, one should not neglect this term. Using $H(a) \simeq \mrm{H}{i}(\mrm{a}{i}/a)^{3(w+1)/2}$ we find the solutions for $w \neq 1$ in the form of
\begin{equation}
\theta(t) = \mrm{\theta}{i} + \frac{2\mrm{\dot{\theta}}{i}}{3(1-w)\mrm{H}{i}}\left [ 1 - \left (\frac{\mrm{a}{i}}{a(t)} \right )^{3(1-w)/2} \right ] \, . \label{eq:axionasymsol}
\end{equation}
We are in particular interested in the solution for radiation domination ($w=1/3$) as a function of scale factor, which reads:
\begin{equation}
	\theta (a) = \mrm{\theta}{i} + \frac{\mrm{\dot{\theta}}{i}}{\mrm{H}{i}}\left ( 1 - \frac{\mrm{a}{i}}{a} \right )\, . \label{eq:axionasymsolrad}
\end{equation}
The above equation is important for understanding the asymptotic behaviour of the axion field after the second reheating as described in the main text. Note that the field values of the solution~$\theta$ may leave the canonical field range of values from $-\otherpi$ to $\otherpi$ as a consequence of~(\ref{eq:axionasymsolrad}). This is fine because the potential energy of the axion only depends on $\cos(\theta)$ and no problems arise if we avoid small-angle approximations.

In the harmonic limit of $\theta \ll 1$ ($\sin (\theta) \simeq \theta$), Equation~(\ref{eq:axionfieldapp}) is a harmonic oscillator with time-dependent damping and mass terms. For $m_A\gg 3H$, which is typically equivalent to the adiabatic limit, the axion field has already started oscillating and one may use the following WKB-inspired ansatz to approximate the solution of~(\ref{eq:axionfieldapp}):
\begin{equation}
\theta (t) = \frac{\theta_\ast}{\cos\(\delta\)} \, \left( \frac{a(t_\ast)}{a(t)} \right)^\frac{1}{2} \, \left( \frac{m_A(t_\ast)}{m_A(t)} \right)^\frac{3}{2} \, \cos \left(\int_{t_\ast}^{t} \! m_A(t) \, \dx t + \delta \right)\, , \label{eq:wkb}
\end{equation}
where we have modified the equation of Arias \textit{et al.}~\cite{1201.5902} by including a phase~$\delta$ to match the analytic ansatz to the numerical result at some time $t_\ast$. The phase~$\delta$ is given by
\begin{equation}
	\delta_{\phantom{\delta}}  = - \arctan \left[\frac{3H_\ast}{2m_\ast} \, \left( 1 + \frac{2\dot{\theta}_\ast}{3\theta_\ast H_\ast} + \frac{\dot{m}_\ast}{3m_\ast H_\ast} \right)\right] \, ,
\end{equation}
where the index ``$\ast$'' refers to evaluation at time~$t_\ast$. One can show the approximate validity of the solution by plugging it back into the field equation. All non-vanishing terms correspond to adiabatic conditions for $H$, $m$ as well as $\dot{m}$ in addition to $3H\ll m_A$ and the harmonic limit of $\theta \ll 1$.

To solve~(\ref{eq:axionfieldapp}), we first solve $3H(T)=m_A(T)$ for the two branches of $H(T)$, obtaining the temperatures $\mrm{T}{osc,1}$ and $\mrm{T}{osc,2}$ for the (nominal) onset of the oscillations before and after instant reheating, respectively. We check for consistency, i.e. $\mrm{T}{osc,1}>\mrm{T}{II}$ and $\mrm{T}{osc,2}<\mrm{T}{II,reh}$. This defines four cases and we solve~(\ref{eq:axionfieldapp}) before and after reheating separately using normalised temperature variables. We also need a stopping condition, which we define as $m_A/3H\geq\alpha_1$ while also $\hat{\theta}\equiv\sqrt{2\rho_A}/f_Am_A\leq\alpha_2$. For our implementation, $\alpha_1=250$ and $\alpha_2=0.01$. Finally, the initial conditions are defined as $\theta\(\alpha_0\)=\mrm{\theta}{i}$ and $\dot{\theta}\(\alpha_0\)=0$.
\begin{enumerate}
	\item Solution before reheating. If $\mrm{T}{osc,1}\leq\mrm{T}{II}$, we solve from $\alpha_0 =\num{e4}\mrm{T}{II}$ until $\mrm{T}{II}$. For $\mrm{T}{osc,1}>\mrm{T}{II}$, however, we first solve from $\alpha_0 = \num{e4}\mrm{T}{osc,1}$ until $\mrm{T}{osc,1}$. We then continue to solve until $\mrm{T}{II}$ \textit{or} until the stopping condition is achieved. In the latter case, we use~(\ref{eq:wkb}) to propagate the field solution to $\mrm{T}{II}$.
	\item Solution after reheating. We use the result from the previous step and demand that the axion solution and its derivative are continuous (as functions of time) at instant reheating. For $\mrm{T}{osc,2}<\mrm{T}{II,reh}$, we solve the field equation first from $\mrm{T}{II,reh}$ until $\mrm{T}{osc,2}$ and then until the stopping condition is achieved. For $\mrm{T}{osc,2}\geq\mrm{T}{II,reh}$, we simply solve the field equation from $\mrm{T}{II,reh}$ until the stopping condition is achieved.
\end{enumerate}
Since we are interested in the axion energy density today, we need to calculate the energy density~$\rho_A$ at the end of our numerical procedure at some temperature~$T_\ast$~\cite[app.~B]{Weinberg:2008zzc}:
\begin{equation}
\rho_A = \frac{1}{2} \, f_A^2 \; \dot{\theta}^2 + V[\theta] \label{eq:axionede} \, .
\end{equation}
In the harmonic and adiabatic limit, one finds that the comoving axion number density is -- on average -- conserved (e.g. by using the WKB-inspired ansatz~\cite{1201.5902}). Once the axion mass becomes constant, the field behaves like cold dark matter. The energy density~(\ref{eq:axionede}) between temperatures $T_\ast$ and $T_o$ (today) scales like
\begin{equation}
\rho_{A,o} = \frac{m_{A,o}}{m_{A,\ast}} \, \frac{g_S(T_o)}{g_S(T_\ast)} \,  \left(\frac{T_o}{T_\ast}\right)^3 \, \rho_{A,\ast} \, , \label{eq:edescaling}
\end{equation}
where the indices refer to the evaluation at the different times, respectively~\cite[p.~430]{Kolb:1990vq}. We use~(\ref{eq:edescaling}) to scale the numerically calculated energy density from the harmonic and adiabatic limit to its value today.

\renewcommand\baselinestretch{0.5}\selectfont
\footnotesize

\bibliographystyle{apsrev_mod}
\bibliography{biblio}

\end{document}